# A NEW SYSTEM FOR RECORDING THE RADIOLOGICAL EFFECTIVE DOSES FOR PACIENTS INVESTIGATED BY IMAGING METHODS

Silviu Stanciu[1,] Lidia Dobrescu[2]
[1] Central Military Emergency Universitary Hospital "Dr. Carol Davila" of Bucharest
[2] University "Politehnica" of Bucharest
silviu.stanciu@yahoo.com,
lidia.dobrescu@electronica.pub.ro

*Abstract-* **In this paper the project of an integrated system for radiation safety and security of the patients investigated by radiological imaging methods is presented. The new system is based on smart cards and Public Key Infrastructure.**
**The new system allows radiation effective dose data storage and a more accurate reporting system.**

## I. INTRODUCTION

In accordance with European Directive 97/43 EURATOM implemented in national regulations, all the ionizing radiation exposures during medical investigations must be carefully recorded and reported.

The International Commission on Radiological Protection (ICRP) recommends that the public limit of artificial irradiation should not exceed an average of 1 mSv effective dose per year, not including medical and occupational exposures. ICRP limits for occupational workers are 20 mSv per year, averaged over defined periods of five years, with the further provision that the dose should not exceed 50 mSv in any single year [1].

The current medical practice uses paper forms for recording different types of investigations and their individual radiation absorbed dose. The patients often sign a formal agreement. Modern radiological apparatus for computerized topographies or scintigraphies can provide the radiation doses during a particular investigation, but the types of doses and their measurement units in different types of investigations are not the same.

The accuracy of the reporting system consists in counting the same type of investigations and the total number is multiplied by a medium dose which can be different from cumulating the total recorded individual doses.

Modern equipment of radiological investigations is continuously reducing the total dose of radiation due to improved films, film screen systems or improved technologies and so there is a substantially decrease in per caput dose, but the increasing number of investigations has determined a net increased annual collective dose.

A secure integrated system is designed. The new system, designed on smart cards technology, covers a major need of the health-care system. Integration of PKI infrastructures is intended to supply a high level of security for the whole system including access to databases through various applications and also to ensure the confidentiality of citizens' personal data stored on cards and in a central data base.

## II. RADIATION DOSES

In Romania, statistics show that three million people are investigated by radiographies and CTs and the investigations by radiological methods strongly increases the cumulative radiation dose of the patients. So, despite the great concern of radiation exposure, the radiological imaging investigations are not very carefully prescribed. The radiographies, CT-s and generally Xrays investigations can save lives but their high level radiation doses can affect people health. Despite the skyrocketing volume of imaging investigations with radiation risk, a major lack in monitoring and track the cumulative radiation doses of the patients that are usually treated and evaluated has become a serious problem and this problem can be solved by a new electronic integrated system based on smartcards.

The most performed examinations in Romania are Chest/Torax, Cervical, Thoracic and Lumbar Spine Radiographies and CT for head, neck, chest, spine and abdomen [2].

The Sievert (Sv) is the International System of Units (SI) derived unit of radiation dose. Confusion can be caused as there is one more radiation unit, the Gray. This last one is used for describing the absorbed dose in any material, while the Sievert is used with effective absorbed dose in biological tissue. So, in order to convert the equivalent dose provided by the radiation source from the medical investigation apparatus, the absorbed dose measured in mGy must be multiplied by a tissue factor which is usually below unit, shown in table I[3].

A difficult situation was determined by CT-s. The dose provided by modern electronic equipments is expressed in CTDI or DLP. Two related measures of CT radiation dose are available on CT consoles: the CT dose index (CTDI) and the dose length product (DLP). The CTDI represents the radiation dose of a single CT slice and is determined using acrylic phantoms. These acrylic phantoms are cylinders of a standard length and are generally in diameters of 16 cm and 32 cm.

The dose length product (DLP) is the $CTDI_{vol}$ multiplied by the scan length (slice thickness × number of slices) in centimeters. It should be noted that the DLP is independent of what is actually scanned. In other words, the reported DLP is the same whether a 10-lb infant or a 100-lb teenager is scanned if the scan length and other scan parameters are the same. DLP was chosen as input data for the system in CTs investigations but the reported data is the effective dose. Conversion factors can be used to estimate what the effective dose equivalent would be. However, these conversion factors are problematic in that they are only estimates of dose and do not represent the full range of pediatric sizes [4]. The conversion factors can slightly vary from different manufacturers [5].

For classic radiological investigation, an important topic for data management is the measurement and the calculation of radiation dose expressed in DAP. Dose area product (DAP) is used to measure the radiation risk from diagnostic x-ray examinations. It is defined as the absorbed dose multiplied by the area irradiated. It is expressed in (Gy*cm2). DAP reflects not only the dose within the radiation field but also the area of tissue irradiated. It is an indicator of the overall risk of inducing cancer. It also has the advantages of being easily measured, with a DAP meter on the x-ray set.

Table I [3].

TISSUE WEIGHTING FACTORS

| Tissue | Tissue weighting factor wT | ΣwT |
|---|---|---|
| Bone-marrow (red), colon, lung, stomach, breast, adrenals, extrathoracic region, gall bladder, heart, kidneys, lymphatic nodes, muscle, oral mucosa, pancreas, prostate small intestine, spleen, thymus, uterus/cervix | 0.12 | 0.72 |
| Gonads | 0.08 | 0.08 |
| Bladder, oesophagus, liver, thyroid | 0.04 | 0.16 |
| Bone surface, brain, salivary glands, skin | 0.01 | 0.04 |
| | Total | 1.00 |

| Measured DAP | 197 mGycm$^2$ |
|---|---|
| Film Area | 1.225 cm2 |
| Skin dose | 0.16 mGy |
| Lung factor | 0.12 |
| Effective dose | 0.0192 mSv |
| | 197 : 1.225 = 0.16 * 0.12 = 0.019 |

Fig1 Effective dose calculus

Due to the divergence of a beam emitted from X ray source, the irradiated area increases with the square of distance from the source, while radiation intensity decreases according to the inverse square of distance.

So DAP as the product of intensity and area become independent of distance from the source. A DAP calculus example is shown in figure 1.

More refinements, shown in table II, can be used in order to consider attributable lifetime risk based upon a relative risk of 1 at age 30 (population average risk). It assumes the multiplicative risk projection model, averaged for the two sexes. In fact, risk for females is always relatively higher than males. Beyond 80 years of age, the risk becomes negligible because the latent period between X-ray exposure and the clinical presentation of a tumor will probably exceed the life span of a patient. In contrast, the tissues of younger people are more radiosensitive and their prospective life span is likely to exceed the latent period.

Radiation is a natural phenomenon in our environment. There are forms of invisible radiation from space or in the earth, in the air, in the water, in our food, all known as natural background radiation. These radiation doses will be added to those from radiological imaging methods of investigation such as computed tomography must become a serious decision in the future. Common CT doses are shown in Table III [6],[7],[8],[9].

Generally speaking ionizing radiation can be harmful and even lethal for humans. In Table IV typically effects are shown. Exposure to ionizing radiation increases the future incidence of cancer, but quantitative models predicting the level of risk are still not worldwide accepted. Induced cancer can be analyzed as a stochastic effect because its probability of occurrence increases with the dose, while

Table II

RISK FACTOR VERSUS AGE

| Age Group (years) | Multiplication factor for risk |
|---|---|
| <10 | x 3 |
| 10-20 | x 2 |
| 20-30 | x 1.5 |
| 30-50 | x 0.5 |
| 50-80 | x 0.3 |
| 80+ | Negligible risk |

the severity is independent of dose, but a threshold dose can be but a threshold dose can be established as in deterministic effects.

### III. CURRENT MEDICAL PRACTICE

During a study in the Central Military Emergency Universitary Hospital Dr. Carol Davila from Bucharest, Romania the patients were monitored for three months. A central data base from this hospital stored patients individual records. Although the hospital's new modern equipment of radiological investigations can provide track information, it is impossible to cumulate all the doses received by a patient. One reason is that many hospitals do not have computerized radiological apparatus. Another reason is that the patients do not have a unique paper form to record all their investigations when they enter in a hospital. Finally, it is not such a difficult practice to repeat a certain investigation in another hospital. The pilot study from Bucharest has revealed many cases of over passing the maximum cumulative dose only during one single hospitalization. The conclusions are shown in figure 2

TABLE III
RADIATION DOSES RECEIVED FROM CT INVESTIGATIONS

| CT examinations | Effective dose (mSv) | Equivalent number of PA chest radiographies |
|---|---|---|
| Head | 2 | 100 |
| Neck | 3 | 150 |
| Calcium scoring | 3 | 150 |
| Pulmonary angiography | 5.2 | 260 |
| Spine | 6 | 300 |
| Chest | 8 | 400 |
| Coronary angiography | 8.7 | 435 |
| Abdomen | 10 | 500 |
| Pelvis | 10 | 500 |
| Virtual colonoscopy | 10 | 500 |
| Chest(pulmonary embolism) | 15 | 750 |

TABLE IV
THRESHOLD RADIATION DOSES FOR HUMAN BODY

| Indicative dose range (mSV) | Effects on human body |
|---|---|
| Up to 10 | No direct evidence on human health effects |
| 10-1000 | No early effects; increased incidence of certain cancers in exposed populations at higher doses |
| 1000-10,000 | Radiation sickness (risk of death); increased incidence of certain cancers in exposed populations |
| Above 10,000 | Always fatal |

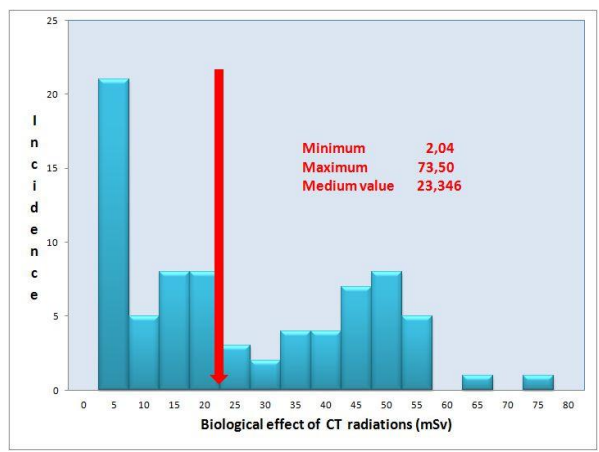

Fig. 2. Cumulative radiation dose in CT

### IV. INTEGRATED SYSTEM REQUIREMENTS AND IMPLEMENTATION

The system provides the replication of the information stored in central databases, local databases and patient cards to cover all the following possible situations:

▪ The patient goes to the doctor without the patient card. In this case, the system provides the data corresponding to the patient based on the local database, if the patient has been investigated in that hospital unit; if the patient is new, the system will provide the data from the central database to the local database, so as to take the optimal decision in recommending the type of investigation. After the investigation, the system will store the new local database accumulated radiation dose to the patient, the information arriving in the central database. Later, when the patient goes to the doctor for further investigations with the card, the system will ensure synchronization between the information stored on the card and information stored in the local and central database.

▪ The patient goes to the doctor with no card and the hospital unit's information system does not have access to the central database (ex: for mobile laboratories). If the local database contains the information corresponding to the patient, the doctor will be able to use them for recommending a particular type of investigation. After the investigation, the system will store in the local database the new radiation doses accumulated by the patient. Later, when the local system can access the central system, the information corresponding to the patient will be synchronized between the two databases.

▪ The patient presents the card to the doctor but the doctor does not have access to any database (local or central). In this case the doctor, using a computer with a card reader can access the history of investigations and the doses received by the patient, as the current cumulative dose calculated and can recommend the most appropriate investigation. After investigation, the appropriate dose will be recorded on the patient's card and next time when the patient goes to the doctor with the card, this information will be stored in both in the local and the central database.

In addition, the system l provides applications for:
▪ Viewing in real time the history of investigations, of the doses delivered to the patient and of the current cumulative calculated dose expressed in mSv.
▪ Aiding the medical staff in taking the adequate decisions regarding the indication of investigations according to the current calculated cumulative doses and the maximum doses allowed for the risk and age groups.
▪ Performing of various periodic reports in order to take different types of decisions related with the existing radioprotection regulations.

The proposed system includes the following components:
- Smart cards dedicated to patients: Citizen Radiation Safety Card – CRSC and Professional Radiation Safety Card – PRSC – intended to medical and investigation laboratories personnel.
- Smart card readers and writers in order to record and retrieve the information about the type of investigations and the specific emitted doses
- A data base records all the necessary information in order to replicate a lost or destroyed card but also this database will provide the possibility of collecting data about the patients on several criteria, it will provide the possibility of standard or customized reports' creation.
- The security solutions such as public key infrastructure PKI in order to achieve a high level of security of recording and retrieving data. A public-key infrastructure (PKI) is a set of hardware, software, people, policies, and procedures needed to create, manage, distribute, use, store, and revoke digital certificates. A PKI establishes and maintains a trustworthy networking environment by providing key and certificate management services that enable encryption and digital signature capabilities across applications — all in a manner that is transparent and easy to use.

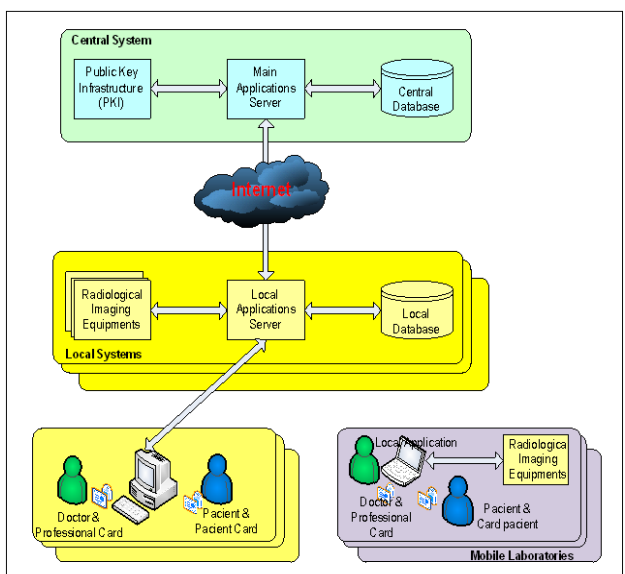

Fig. 3. The architecture of the integrated system.

The system's general architecture has the main components:
▪ Public Key infrastructure used for electronic smartcards operation;
▪ The center for personalising cards where the cards will be created;
▪ The system for the management of the digitized certificates E-HEALTH CMS;
▪ The servers for local databases and central database.
▪ Applications and web services for the interface with the integrated system, electronic cards and medical equipements;
▪ On card Applications called JavaCard applets.

The project consortium has developed a system that integrates all activities with risks of radiations, not only the patients investigated by radiological imaging methods. The smart cards allow authentication, digital signature and secure data storage.

The whole system is designed on two types of radiation safety cards:
- Citizen Radiation Safety Card – CRSC
- Professional Radiation Safety Card – PRSC – for medical and investigation laboratories personnel.

ACKNOWLEDGMENT

The results were obtained from the SRSPIRIM project in Romanian Partnerships Program, Collaborative Applied Research Projects Subprogram. The authors wishes to address thanks to all the persons involved in this project for their support, work, and ideas.